\documentclass{article}
\usepackage{graphicx} 
\usepackage{amsmath}
\usepackage{appendix}
\usepackage{hyperref}

\def\Address{$^{1}$Eindhoven University of Technology, Department of Electrical Engineering, Eindhoven, the Netherlands\\
$^{2}$Epilepsy Center Kempenhaeghe, Department of Behavioral Sciences, Heeze, the Netherlands \\
$^{3}$Philips Healthcare, Image Guided Therapy, Best, the Netherlands}
\def\corrEmail{$^{*}$Correspondence: s.j.c.schielen@tue.nl}

\begin{document}

\title{ICA-based Resting-State Networks Obtained on Large Autism fMRI Dataset ABIDE} 
\date{}
\author{Sjir J.C. Schielen\,$^{1,*}$, Jesper Pilmeyer\,$^{1}$, Albert P. Aldenkamp\,$^{1,2}$, \\ Danny Ruijters\,$^{1,3}$, and Svitlana Zinger\,$^{1}$}
\def\Address{\noindent $^{1}$Eindhoven University of Technology, Department of Electrical Engineering, Eindhoven, the Netherlands\\
$^{2}$Epilepsy Center Kempenhaeghe, Department of Behavioral Sciences, Heeze, the Netherlands \\
$^{3}$Philips Healthcare, Image Guided Therapy, Best, the Netherlands}
\maketitle
\Address \\
\corrEmail

\section{Introduction}

In the exploration of potential differences between people with autism spectrum disorder (ASD) and healthy controls, analyses progressing from functional magnetic resonance imaging (fMRI) have gained popularity \cite{santana2022rs, schielen2024diagnosis}. These studies seek to find a neural signature of ASD by analyzing the blood-oxygen-level-dependent (BOLD) signal, a measure that reflects the concentration of deoxygenated hemoglobin resulting from neural activity. As this signal is measured per voxel, the brain is typically parcellated into regions of interest (ROIs) to reduce dimensionality.

Commonly, this parcellation follows a predefined brain atlas. The popularity of dividing the brain according to a predefined atlas can partly be explained by the easy accessibility of data through the Autism Brain Imaging Data Exchange (ABIDE), which also offers already preprocessed data parcellated with various atlases \cite{di2014autism, di2017enhancing, craddock2013neuro}.

Another option for dimensionality reduction is independent component analysis (ICA), a data-driven form of blind source separation that allows the decomposition of brain fMRI into spatially non-overlapping maps and their time series \cite{calhoun2009review}. These maps either contain noise or patterns of activity scattered over the brain. As these patterns are observable without exerting external stimuli (i.e. task-free), they are called resting-state networks (RSNs) \cite{niazy2015resting, damoiseaux2006consistent}. 

While less frequently explored in the literature, ICA offers some unique benefits. Where atlas-based parcellation segments the brain based on a standard template, ICA derives components directly from the study's data, thereby tailoring the delineation to the studied sample. Additionally, this happens without hypotheses on spatial regions, allowing neural activation to be clustered without presumed delineations \cite{calhoun2009review}. Therefore, we have preprocessed and used group ICA on 900 individuals from ABIDE to identify group RSNs, from which subject-specific time series were obtained using dual regression \cite{beckmann2009group}, \cite{nickerson2017using}. This data is publicly available to complement ABIDE's repository of preprocessed data. With this alternative approach to atlas-based parcellation, we contribute to the research that seeks to understand the complex neural dynamics of ASD.

\section{Methods}

The methods section is divided into two parts. The first part outlines the steps involved in selecting data from ABIDE. The second part contains descriptions of the preprocessing steps.  

\subsection{Data Selection}

\begin{figure}
    \centering
    \includegraphics[width=\textwidth]{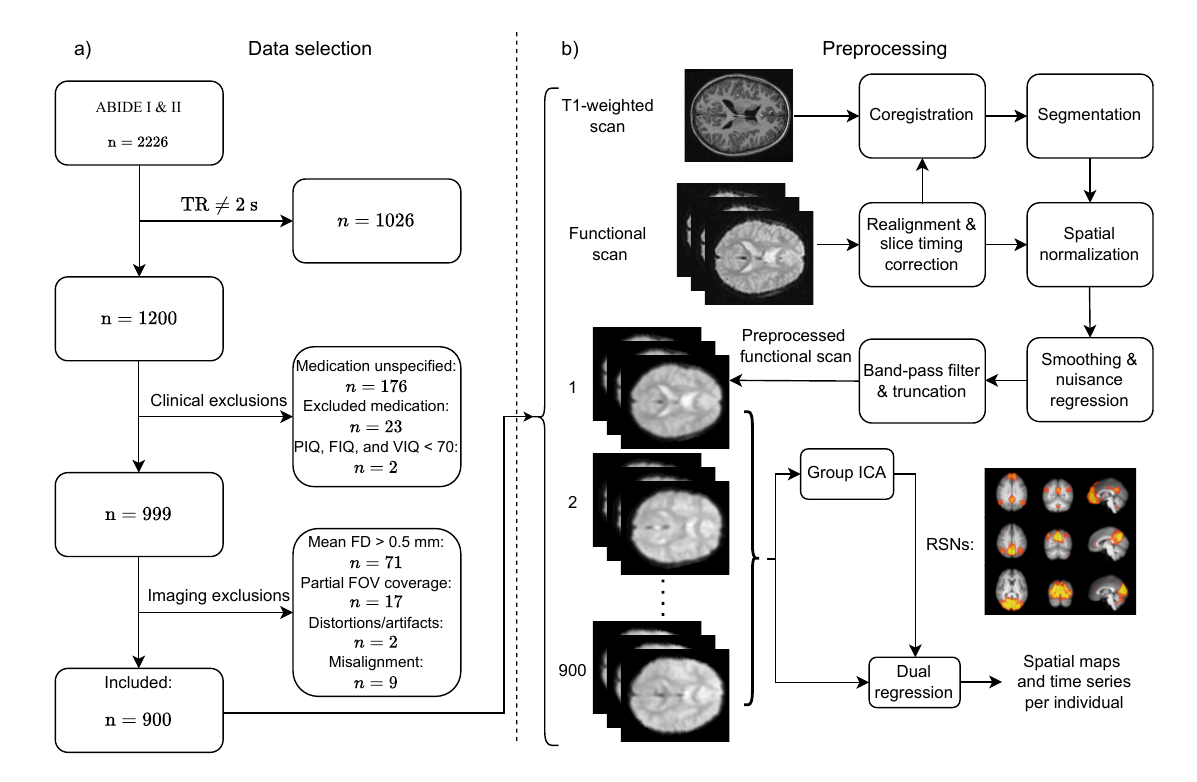}
    \caption{The schematic presentation of the selection process (a) and the preprocessing steps (b), where $n$ is used to indicate the sample size, $\text{TR}$ the repetition time, \{P, F, V\} IQ respectively the \{Performance, Full-scale, Verbal\} Intelligence Quotient, FD framewise displacement, FOV the field of view, ICA independent component analysis, and RSNs resting-state networks. }
    \label{fig:selectionprep}
\end{figure}

To ensure the dataset was suitable for group ICA and comparisons between individuals, we made the selections summarized in Fig. \ref{fig:selectionprep}a. Initially, we considered all cross-sectional data from ABIDE. As the dataset is preprocessed for group ICA, it requires consistent repetition times (TRs) across subjects because signals are temporally concatenated \cite{beckmann2004probabilistic}. Thus, subjects scanned with a TR other than 2000 ms (the most common) were excluded (n=1026). While temporal interpolation could have allowed the inclusion of more subjects, we prioritized maintaining signal integrity where possible. 

Then, potential sources of clinical heterogeneity were considered. Although there is some variation throughout the ABIDE sample in diagnostic assessment and instructions for the resting task (e.g. eyes open or closed), we deemed the potential effects of these variations negligible given the sample size. We recognize that psychoactive medication can significantly affect brain function \cite{linke2017psychotropic}. Consequently, all patients on centrally active medication potentially affecting fMRI (i.e. valproic acid, oxcarbazepine, topiramate, risperidone, citalopram, and lamotrigine) and the sites that did not report this information were excluded (n=23 and n=176, respectively).

Although comorbidities can obscure the interpretation of fMRI results, their high prevalence in ASD led us not to exclude comorbidities \cite{bartolotti2020functional, antshel2016update, lai2019prevalence}. Specifically, 59\% of the autistic population in the sample of ABIDE-II has at least one comorbidity \cite{di2017enhancing}. Rather than excluding comorbidities, we only excluded subjects for which the performance, full-scale, and verbal intelligence quotient (\{P, F, V\} IQ) were all lower than 70 (n=2). As these cases were rare in the sample, participants were not excluded if any of the IQ types were not reported.   

After the clinical exclusions, 999 subjects were included for preprocessing. As motion can degrade signal quality and introduce spurious findings in fMRI \cite{power2012spurious}, subjects with a mean framewise displacement over 0.5 mm were excluded (n=71). Furthermore, exclusions based on artifacts were made if the issue caused inappropriate comparisons between individuals. In some scans (n=17), brain parts of interest (often the cerebellum, but occasionally parts of the occipital or temporal lobe) fell outside the field of view (FOV). While some artifacts are rather common (e.g. intensity non-uniformity or susceptibility), exclusions (n=2) were only made if the artifact changed over time or caused signal loss in areas of interest. When the scans were brought into the standard reference space \cite{evans19933d}, exclusions (n=9) were made if the brain significantly deviated from alignment with other brains.

Finally, 900 individuals were included from the ABIDE-I sites: Carnegie Mellon University (CMU), University of Michigan (UM; two samples), and Yale Child Study Center (Yale); the ABIDE-II sites: Erasmus University Medical Center Rotterdam (EMC), Georgetown University (GU), University of California Davis (UCD), University of Miami (UMia), and University of Utah School of Medicine (USM); and the sites both in ABIDE-I and ABIDE-II: New York University Langone Medical Center (NYU; both ABIDE-II samples), San Diego State University (SDSU), Stanford University (SU), and Trinity Centre for Health Services (TCD). As some sites are in both iterations of ABIDE, site abbreviations contain the prefix A1 or A2 to respectively represent the sample from ABIDE-I and ABIDE-II when necessary.

\subsection{Preprocessing}
Fig. \ref{fig:selectionprep}b illustrates the step-by-step preprocessing approach aimed at preparing functional scans for group ICA. Then, dual regression was used to obtain spatial maps and time series per individual. Each step is detailed below. The preprocessing approach and scripts were based on \cite{heunis2021fmriwhy}.

Realignment primarily minimizes misalignment between subsequent scanned volumes via image registration. After discarding the first four volumes of each functional scan to compensate for non-steady-state signals, each volume in the scan was registered to the first using the MCFLIRT function of FSL \cite{jenkinson2002improved}. Image registrations were performed with six degrees of freedom (motion parameters): three for rotation and three for translation.

The summation of the absolute difference between motion parameters per time step, i.e. the framewise displacement (FD), was used to exclude subjects if the mean FD exceeded 0.5 mm. While it is argued that even movements of small magnitude (less than 0.05 mm) can have artifactual effects on fMRI data \cite{byrge2018identifying}, motion is common in the dataset and there is no consensus on quality assessment standards \cite{di2017enhancing}. Therefore, fMRI-based findings, e.g. biomarkers, should be robust to some extent of motion, which is decided to be up to a mean framewise displacement of 0.5 mm here. 

Slice timing correction temporally interpolates slices to account for acquisition times within a single volume. As this dataset is intended for methodologies that investigate differences between ASD and controls (e.g. group differences, diagnostic tools, or biomarker identification for ASD), the preprocessing is tailored to accommodate multiple potential analyses. Common methods are analyses progressing from functional or effective connectivity \cite{friston2011functional}, where the latter estimates a sense of causality that requires correct timing. E.g., dynamic causal modelling is a method of estimating effective connectivity, which requires slice timing correction \cite{friston2003dynamic}. Moreover, different acquisition protocols were used (sequential and interleaved slice acquisitions), which means comparisons between scans acquired with different slice orders are less appropriate than when corrected. Slice timing correction was performed referencing the middle slice using SPM. 

Coregistration aligns the anatomical scan with the functional scan, leveraging the anatomical scan's higher spatial resolution for segmentation. Coregistration was performed using SPM with default settings. For some subjects, the spatial alignment between the functional and anatomical scans differed significantly which caused coregistration to fail. In these cases, a manual translation was performed to create a better starting position for the registration. If the coregistration failed after this effort and the misalignment caused incorrect spatial normalization, then the subject was excluded (n=9).

Segmentation was performed in SPM with default settings to obtain probability maps of air, soft tissue, bone, grey matter, white matter, and cerebrospinal fluid in the anatomical scan. These maps were later used for masking. In the segmentation step, the forward transformation of native space to Montreal Neurological Institute (MNI) space was also obtained \cite{evans19933d}. The forward transformation was applied to the functional image in the spatial normalization step, which also involved 4th-order B-spline interpolation to make the spatial resolution 2x2x2 mm. Normalization to the same space and a common voxel size among subjects scanned with different acquisition protocols was necessary to ensure correct spatial comparison between subjects. 

Smoothing is commonly used in fMRI preprocessing pipelines to increase the signal-to-noise ratio and lower inter-subject variability \cite{blazejewska2019intracortical}. SPM was used to perform smoothing with a three-dimensional Gaussian kernel with a full width at half maximum of 5 mm in each dimension. Despite smoothing, realignment, and exclusions of excessive FD, secondary effects of motion show up artifactually in the fMRI signal which may result in spurious findings of functional connectivity \cite{balachandrasekaran2021reducing, power2012spurious}. Therefore, ICA-based Automatic Removal of Motion Artifacts (ICA-AROMA) was used to decompose the fMRI signal and automatically classify and regress out motion-related noise components \cite{pruim2015ica}. We chose ICA-AROMA for its effectiveness in removing motion-induced correlations between distant voxels while preserving functional network identifiability \cite{ciric2017benchmarking}. Extending ICA-AROMA with global signal regression (GSR) could eliminate more noise components from the signal at the risk of losing signal valuable to e.g. functional network identifiability \cite{murphy2017towards}. Considering the manual selection of components was performed after group ICA, we decided not to extend ICA-AROMA with GSR as only the components corresponding to RSNs were selected. 

A band-pass filter passing 0.01 to 0.1 Hz was applied to restrict the fMRI signal to a neural frequency range while minimizing cardiac and respiratory interference \cite{margulies2010resting, boubela2013beyond}. This was implemented with a second-order zero-phase digital Butterworth filter in Matlab R2023b. To ensure equal weighting between individuals in group ICA, all fMRI scans were truncated to 146 volumes, corresponding to the shortest length of the included scans minus the four discarded volumes at the start. The truncated volumes all contain the first 146 volumes (after discarding), as exclusion-worthy artifacts or distortions were mostly observed later on in the scans.

Group ICA was performed using FSL's Multivariate Exploratory Linear Optimized Decomposition into Independent Components (MELODIC). Preprocessed functional scans were temporally concatenated and the implementation of probabilistic ICA was used to perform the decomposition \cite{beckmann2004probabilistic}. Increasing the model order (i.e. the number of components) tends to decompose networks further into sub-networks \cite{abou2010effect}. Therefore, multiple iterations of this process were run, each with a different number of components: 30, 32, 50, and 'default' in which the number of components was estimated automatically to be 74. The components corresponding to RSNs were manually selected. 

Dual regression was used to calculate each individual's version of the group-level components and their associated time series \cite{filippini2009distinct}. The benefit of dual regression is that the group ICA components serve as the design matrix in the general linear model, which is optimized to resemble the individual's preprocessed functional scan. This captures the individual's variability while preserving the spatial characteristics of the group ICA.

\section{Data Records}
In this section, an overview of the available data is provided. All data can be found on the GitHub page: https://github.com/SjirSchielen/groupICAonABIDE. 

\subsection{Resting-state Networks}
\begin{figure}[ht]
    \centering
    \includegraphics[width=\textwidth]{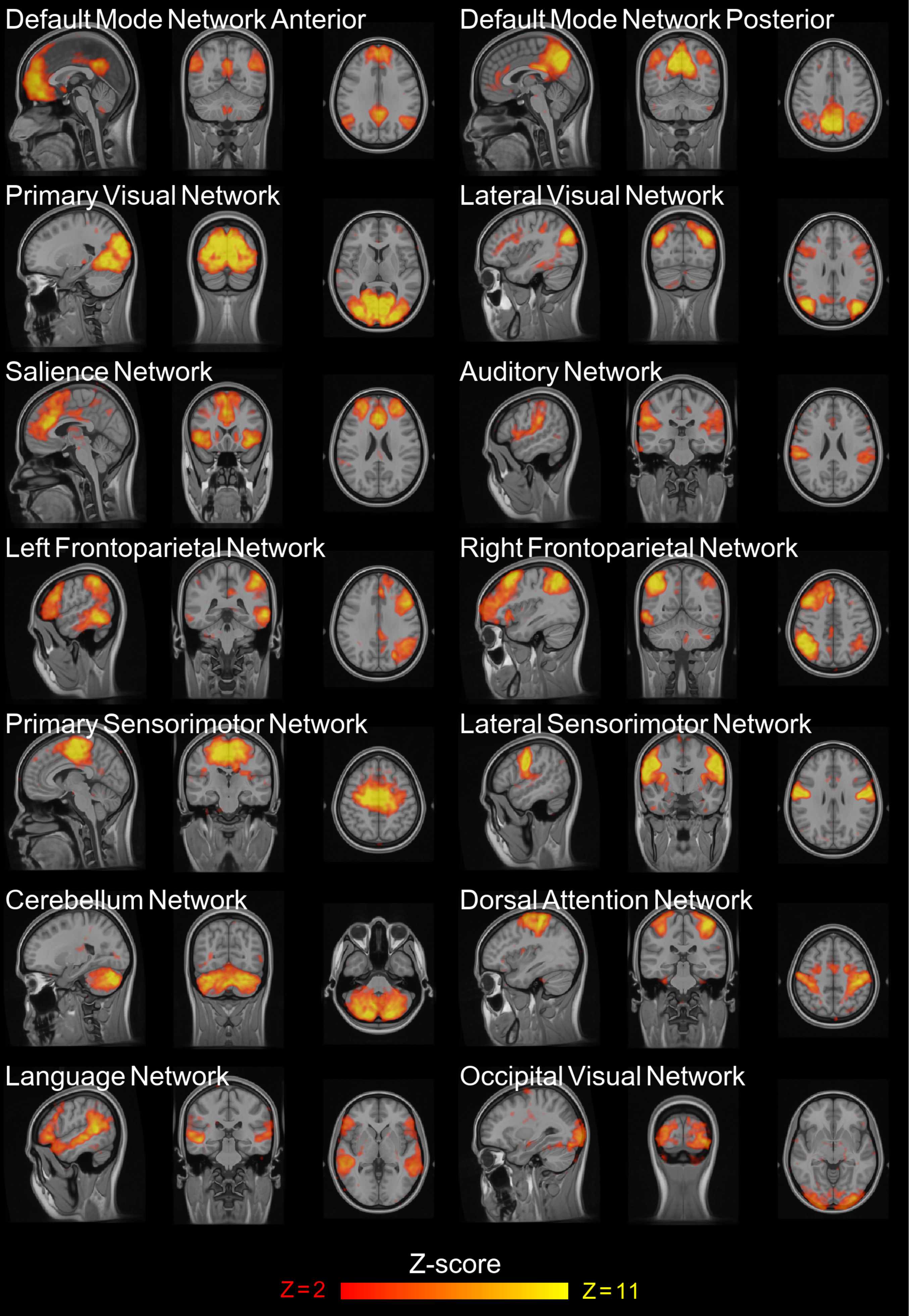}
    \caption{The group independent components that were identified to correspond to resting-state networks, where the Z-score refers to the number of standard deviations a voxel's value is from the mean.}
    \label{fig:RSNs}
\end{figure}

The group ICA components identified as resting-state networks (RSNs) are overlaid with the MNI 152 ICBM template image \cite{fonov2009unbiased}, as shown in Fig. \ref{fig:RSNs}. As described in the literature \cite{abou2010effect}, increasing the number of components led to networks being split into sub-networks. While using 50 or 74 components split most networks into sub-networks, using 30 components resulted in networks that were not sufficiently separated from noise. The 32-component ICA showed components separated from noise while only a few networks were split into sub-networks. Therefore, components were selected from the 32-component ICA. This identification process was a collaborative effort among the authors, utilizing the Smith functional brain atlas as an initial reference \cite{smith2009correspondence}. Given the variability in naming conventions for RSNs across studies \cite{uddin2019towards}, we have proposed names for the networks in Fig. \ref{fig:RSNs} that align with prevalent terminology in the field. To foster research flexibility, we have made the output components publicly available on GitHub, allowing users to select components that best suit their research interests and the names they are familiar with.

\subsection{Phenotypic and Demographic Information}
A phenotypic overview of the dataset is listed in Table S1. The dataset contains 417 individuals with ASD (361 male, 56 female) and 483 healthy controls (377 male, 106 female). The ASD group has a mean age of 12.84 years ($\pm$ a standard deviation of 5.04) and the control group has a mean age of 13.84 years ($\pm$ 5.20). The average PIQ of the ASD group is lower 105.78 ($\pm$ 17.31) than the control group  108.94 ($\pm$ 14.54). While the dataset reflects the commonly observed male predominance in ASD diagnoses \cite{zeidan2022global}, it also presents an opportunity to explore the less represented female perspective in ASD research. Although the ASD and HC groups show statistical differences in terms of sex (P=0.0117, $\chi^2$=11.00, Chi-Square test), age (P=0.0017, U=88519, Mann-Whitney U test), and Performance IQ (P=0.0040, t=-2.89, Welch’s test), these variations are reflective of the broader ASD population \cite{zeidan2022global}. To accommodate studies requiring matched cohorts, the dataset allows for the selection of subsets. 

Site distributions of diagnosis, sex, age, PIQ, and mean FDs are shown in supplementary Fig. S1a to S1e. The occurrences of comorbidities in the dataset are shown in supplementary Fig. S1f. Note that the comorbidities in Fig. S1f do not sum to the total number of people with comorbidities, as 32 of the 104 people with comorbidities in the dataset have two or more. All phenotypic information in ABIDE is also available in this dataset on GitHub. In addition, there is a Python notebook (phenotypicDataLoader.ipynb) that acts as sample code to load the phenotypic data, allowing easy use of the phenotypic data for further analyses or data selection.

\section{Analysis}
In this section, data analysis in the form of validation and permutation testing are described. 
\subsection{Validation}
In the ABIDE-II initiative, data was shared regardless of imaging quality because of the absence of a consensus on quality criteria and to accommodate the development of artifact correction methods \cite{di2017enhancing}. The dataset we present here was selected to perform group ICA from which further analyses can progress. Therefore, care was taken to ensure proper comparisons between subjects in the dataset. 

To validate that preprocessed functional scans can be compared properly, we visually inspected all 999 preprocessed functional scans in the dataset and discussed potential exclusions until a consensus was reached among the authors. We prioritized two main points: spatial alignment should be correct to ensure the same regions are considered among participants in ICA and artifacts or distortions should not change over time which might cause spurious findings of activation patterns. 

In most cases in which artifacts changed over time, the mean FD was higher than 0.5 mm for which they were excluded. After visual inspection, two more cases were excluded for time-varying artifacts: subject identifier (SID) 28901 of A2-SDSU and 51176 of A1-SU.

When important brain parts fall outside of the FOV, their signals are not measured. For example, for  50653 (A1-CMU) the field of view missed parts of the temporal lobe and cerebellum. Another example is SID 51167 (A1-SU) in which parts of the parietal lobe were out of view and parts of the occipital lobe wrapped around. Finally, not all coregistrations were successful which was judged by visual misalignment to the reference space. Not all artifacts were excluded. Types of artifacts that are still present in the dataset are both darker (e.g. SID 50958 A1-NYU) and brighter (e.g. SID 28755 A2-GU) susceptibility artifacts, intensity non-uniformity (e.g. SID 50603 A1-Yale), and slight wrap-around artifacts (e.g. SID 28755 A2-GU). Artifacts are also part of a realistic dataset, so subsequent analysis should have the opportunity to incorporate some robustness. 

\subsection{Permutation testing}
While the group ICA components and the subject-specific time series are the main contributions of this paper, further analysis was performed through randomized permutation testing. The second step in dual regression involves regressing the subject-specific time series per RSN (as temporal regressors in a multiple regression) into the subject's preprocessed functional scan, resulting in subject-specific versions of each group-level spatial map. Each subject-specific RSN was then used in FSL's randomise permutation-testing tool to perform a two-sample unpaired t-test for differences between the ASD and control group while accounting for the nuisance variables age and sex \cite{nickerson2017using}.

No significant differences were found with permutation testing. The lowest P-value (P=0.078) was found for the occipital visual network. The full results of permutation testing are listed in Table S2. The lack of significant P-values shows that there are no significant differences in the subject-specific spatial maps between subjects with ASD and controls. As the subject-specific spatial maps result from regressing the subject-specific time series into a subject's 4D scan, there is no significant difference in which voxels contribute to this temporal behavior between ASD and control for each RSN. Therefore, no significant structural differences in the RSNs were found between ASD and control. While the subject-specific spatial maps result from regression involving time series, they summarize this information in one 3D volume. As there is no significant structural difference in the RSNs between ASD and control, it merits further temporal analysis.    

\section{Limitations}
The readily available dataset is based on group ICA, which inherently limits the generalizability of results to individuals not included in the group. Consequently, the presented dataset can be viewed as a case study, e.g. for diagnostic purposes, which requires adopting the same selection criteria, while allowing for further personalized selections. 
However, the identified resting-state networks can be extended to individuals outside the sample by using the group ICA components as the design matrix in dual regression. Since the group ICA components are purely spatial, they can be applied to individuals scanned with different repetition times, similar to using a parcellation atlas. This flexibility enables adaptations of the methodology to various datasets, although it remains a limitation that those individuals were not included in the original group ICA.

\section*{Conflict of Interest Statement}
We declare there are no competing interests.

\section*{Author Contributions}

S.J.C.S.: conception, preprocessing, visual inspection, ICA component selection, manuscript. J.P.: preprocessing, visual inspection, ICA component selection. A.P.A.: patient selection, ICA component selection, supervision. D.R.: preprocessing, supervision. S.Z.: conception, preprocessing, ICA component selection, supervision. All authors reviewed the manuscript. 

\section*{Funding}
This work is only funded by Eindhoven University of Technology, department of Electrical Engineering. 

\section*{Acknowledgments}
We acknowledge the significant contributions of ABIDE-I and ABIDE-II, their primary sources of funding: NIMH K23MH087770 and NIMH 5R21MH107045, respectively \cite{di2014autism, di2017enhancing}, and the secondary sources of funding not listed here.

\section*{Data Availability Statement}
The preprocessing code can be found at \\https://github.com/SjirSchielen/groupICAonABIDE/tree/main/code. Note that this pipeline was tailored to Matlab R2023b on Windows, which requires SPM12 \cite{penny2011statistical} and the image processing toolbox. Furthermore, FSL 6.0.7.6 \cite{jenkinson2012fsl} and the scripts and resources for ICA-AROMA \cite{pruim2015ica} were used. FSL was run in the Windows Subsystem for Linux (WSL), which involved adapting commands and the ICA-AROMA scripts to work in WSL. These versions are available in the repository. 

\bibliographystyle{unsrt}
\bibliography{bib}

\begin{thebibliography}{10}

\bibitem{santana2022rs}
Caio~Pinheiro Santana, Emerson~Assis de~Carvalho, Igor~Duarte Rodrigues, Guilherme~Sousa Bastos, Adler~Diniz de~Souza, and Lucelmo~Lacerda de~Brito.
\newblock rs-fmri and machine learning for asd diagnosis: A systematic review and meta-analysis.
\newblock {\em Scientific reports}, 12(1):6030, 2022.

\bibitem{schielen2024diagnosis}
Sjir~JC Schielen, Jesper Pilmeyer, Albert~P Aldenkamp, and Svitlana Zinger.
\newblock The diagnosis of asd with mri: a systematic review and meta-analysis.
\newblock {\em Translational Psychiatry}, 14(1):318, 2024.

\bibitem{di2014autism}
Adriana Di~Martino, Chao-Gan Yan, Qingyang Li, Erin Denio, Francisco~X Castellanos, Kaat Alaerts, Jeffrey~S Anderson, Michal Assaf, Susan~Y Bookheimer, Mirella Dapretto, et~al.
\newblock The autism brain imaging data exchange: towards a large-scale evaluation of the intrinsic brain architecture in autism.
\newblock {\em Molecular psychiatry}, 19(6):659--667, 2014.

\bibitem{di2017enhancing}
Adriana Di~Martino, David O’connor, Bosi Chen, Kaat Alaerts, Jeffrey~S Anderson, Michal Assaf, Joshua~H Balsters, Leslie Baxter, Anita Beggiato, Sylvie Bernaerts, et~al.
\newblock Enhancing studies of the connectome in autism using the autism brain imaging data exchange ii.
\newblock {\em Scientific data}, 4(1):1--15, 2017.

\bibitem{craddock2013neuro}
Cameron Craddock, Yassine Benhajali, Carlton Chu, Francois Chouinard, Alan Evans, Andr{\'a}s Jakab, Budhachandra~Singh Khundrakpam, John~David Lewis, Qingyang Li, Michael Milham, et~al.
\newblock The neuro bureau preprocessing initiative: open sharing of preprocessed neuroimaging data and derivatives.
\newblock {\em Frontiers in Neuroinformatics}, 7(27):5, 2013.

\bibitem{calhoun2009review}
Vince~D Calhoun, Jingyu Liu, and T{\"u}lay Adal{\i}.
\newblock A review of group ica for fmri data and ica for joint inference of imaging, genetic, and erp data.
\newblock {\em Neuroimage}, 45(1):S163--S172, 2009.

\bibitem{niazy2015resting}
Rami~K Niazy, David~M Cole, Christian~F Beckmann, and Stephen~M Smith.
\newblock Resting-state networks.
\newblock {\em fMRI: From Nuclear Spins to Brain Functions}, pages 387--425, 2015.

\bibitem{damoiseaux2006consistent}
Jessica~S Damoiseaux, Serge~ARB Rombouts, Frederik Barkhof, Philip Scheltens, Cornelis~J Stam, Stephen~M Smith, and Christian~F Beckmann.
\newblock Consistent resting-state networks across healthy subjects.
\newblock {\em Proceedings of the national academy of sciences}, 103(37):13848--13853, 2006.

\bibitem{beckmann2009group}
Christian~F Beckmann, Clare~E Mackay, Nicola Filippini, Stephen~M Smith, et~al.
\newblock Group comparison of resting-state fmri data using multi-subject ica and dual regression.
\newblock {\em Neuroimage}, 47(Suppl 1):S148, 2009.

\bibitem{nickerson2017using}
Lisa~D Nickerson, Stephen~M Smith, D{\"o}st {\"O}ng{\"u}r, and Christian~F Beckmann.
\newblock Using dual regression to investigate network shape and amplitude in functional connectivity analyses.
\newblock {\em Frontiers in neuroscience}, 11:115, 2017.

\bibitem{beckmann2004probabilistic}
Christian~F Beckmann and Stephen~M Smith.
\newblock Probabilistic independent component analysis for functional magnetic resonance imaging.
\newblock {\em IEEE transactions on medical imaging}, 23(2):137--152, 2004.

\bibitem{linke2017psychotropic}
Annika~C Linke, Lindsay Olson, Yangfeifei Gao, Inna Fishman, and Ralph-Axel M{\"u}ller.
\newblock Psychotropic medication use in autism spectrum disorders may affect functional brain connectivity.
\newblock {\em Biological Psychiatry: Cognitive Neuroscience and Neuroimaging}, 2(6):518--527, 2017.

\bibitem{bartolotti2020functional}
James Bartolotti, John~A Sweeney, and Matthew~W Mosconi.
\newblock Functional brain abnormalities associated with comorbid anxiety in autism spectrum disorder.
\newblock {\em Development and psychopathology}, 32(4):1273--1286, 2020.

\bibitem{antshel2016update}
Kevin~M Antshel, Yanli Zhang-James, Kayla~E Wagner, Ana Ledesma, and Stephen~V Faraone.
\newblock An update on the comorbidity of adhd and asd: A focus on clinical management.
\newblock {\em Expert review of neurotherapeutics}, 16(3):279--293, 2016.

\bibitem{lai2019prevalence}
Meng-Chuan Lai, Caroline Kassee, Richard Besney, Sarah Bonato, Laura Hull, William Mandy, Peter Szatmari, and Stephanie~H Ameis.
\newblock Prevalence of co-occurring mental health diagnoses in the autism population: a systematic review and meta-analysis.
\newblock {\em The Lancet Psychiatry}, 6(10):819--829, 2019.

\bibitem{power2012spurious}
Jonathan~D Power, Kelly~A Barnes, Abraham~Z Snyder, Bradley~L Schlaggar, and Steven~E Petersen.
\newblock Spurious but systematic correlations in functional connectivity mri networks arise from subject motion.
\newblock {\em Neuroimage}, 59(3):2142--2154, 2012.

\bibitem{evans19933d}
Alan~C Evans, D~Louis Collins, SR~Mills, Edward~D Brown, Ryan~L Kelly, and Terry~M Peters.
\newblock 3d statistical neuroanatomical models from 305 mri volumes.
\newblock In {\em 1993 IEEE conference record nuclear science symposium and medical imaging conference}, pages 1813--1817. IEEE, 1993.

\bibitem{heunis2021fmriwhy}
Stephan Heunis.
\newblock jsheunis/fmrwhy: Release 0.0.2 - for referencing, 2021.

\bibitem{jenkinson2002improved}
Mark Jenkinson, Peter Bannister, Michael Brady, and Stephen Smith.
\newblock Improved optimization for the robust and accurate linear registration and motion correction of brain images.
\newblock {\em Neuroimage}, 17(2):825--841, 2002.

\bibitem{byrge2018identifying}
Lisa Byrge and Daniel~P Kennedy.
\newblock Identifying and characterizing systematic temporally-lagged bold artifacts.
\newblock {\em NeuroImage}, 171:376--392, 2018.

\bibitem{friston2011functional}
Karl~J Friston.
\newblock Functional and effective connectivity: a review.
\newblock {\em Brain connectivity}, 1(1):13--36, 2011.

\bibitem{friston2003dynamic}
Karl~J Friston, Lee Harrison, and Will Penny.
\newblock Dynamic causal modelling.
\newblock {\em Neuroimage}, 19(4):1273--1302, 2003.

\bibitem{blazejewska2019intracortical}
Anna~I Blazejewska, Bruce Fischl, Lawrence~L Wald, and Jonathan~R Polimeni.
\newblock Intracortical smoothing of small-voxel fmri data can provide increased detection power without spatial resolution losses compared to conventional large-voxel fmri data.
\newblock {\em NeuroImage}, 189:601--614, 2019.

\bibitem{balachandrasekaran2021reducing}
Arvind Balachandrasekaran, Alexander~L Cohen, Onur Afacan, Simon~K Warfield, and Ali Gholipour.
\newblock Reducing the effects of motion artifacts in fmri: A structured matrix completion approach.
\newblock {\em IEEE transactions on medical imaging}, 41(1):172--185, 2021.

\bibitem{pruim2015ica}
Raimon~HR Pruim, Maarten Mennes, Daan van Rooij, Alberto Llera, Jan~K Buitelaar, and Christian~F Beckmann.
\newblock Ica-aroma: A robust ica-based strategy for removing motion artifacts from fmri data.
\newblock {\em Neuroimage}, 112:267--277, 2015.

\bibitem{ciric2017benchmarking}
Rastko Ciric, Daniel~H Wolf, Jonathan~D Power, David~R Roalf, Graham~L Baum, Kosha Ruparel, Russell~T Shinohara, Mark~A Elliott, Simon~B Eickhoff, Christos Davatzikos, et~al.
\newblock Benchmarking of participant-level confound regression strategies for the control of motion artifact in studies of functional connectivity.
\newblock {\em Neuroimage}, 154:174--187, 2017.

\bibitem{murphy2017towards}
Kevin Murphy and Michael~D Fox.
\newblock Towards a consensus regarding global signal regression for resting state functional connectivity mri.
\newblock {\em Neuroimage}, 154:169--173, 2017.

\bibitem{margulies2010resting}
Daniel~S Margulies, Joachim B{\"o}ttger, Xiangyu Long, Yating Lv, Clare Kelly, Alexander Sch{\"a}fer, Dirk Goldhahn, Alexander Abbushi, Michael~P Milham, Gabriele Lohmann, et~al.
\newblock Resting developments: a review of fmri post-processing methodologies for spontaneous brain activity.
\newblock {\em Magnetic Resonance Materials in Physics, Biology and Medicine}, 23:289--307, 2010.

\bibitem{boubela2013beyond}
Roland~N Boubela, Klaudius Kalcher, Wolfgang Huf, Claudia Kronnerwetter, Peter Filzmoser, and Ewald Moser.
\newblock Beyond noise: using temporal ica to extract meaningful information from high-frequency fmri signal fluctuations during rest.
\newblock {\em Frontiers in human neuroscience}, 7:168, 2013.

\bibitem{abou2010effect}
Ahmed Abou-Elseoud, Tuomo Starck, Jukka Remes, Juha Nikkinen, Osmo Tervonen, and Vesa Kiviniemi.
\newblock The effect of model order selection in group pica.
\newblock {\em Human brain mapping}, 31(8):1207--1216, 2010.

\bibitem{filippini2009distinct}
Nicola Filippini, Bradley~J MacIntosh, Morgan~G Hough, Guy~M Goodwin, Giovanni~B Frisoni, Stephen~M Smith, Paul~M Matthews, Christian~F Beckmann, and Clare~E Mackay.
\newblock Distinct patterns of brain activity in young carriers of the apoe-$\varepsilon$4 allele.
\newblock {\em Proceedings of the National Academy of Sciences}, 106(17):7209--7214, 2009.

\bibitem{fonov2009unbiased}
Vladimir~S Fonov, Alan~C Evans, Robert~C McKinstry, C~Robert Almli, and DL~Collins.
\newblock Unbiased nonlinear average age-appropriate brain templates from birth to adulthood.
\newblock {\em NeuroImage}, 47:S102, 2009.

\bibitem{smith2009correspondence}
Stephen~M Smith, Peter~T Fox, Karla~L Miller, David~C Glahn, P~Mickle Fox, Clare~E Mackay, Nicola Filippini, Kate~E Watkins, Roberto Toro, Angela~R Laird, et~al.
\newblock Correspondence of the brain's functional architecture during activation and rest.
\newblock {\em Proceedings of the national academy of sciences}, 106(31):13040--13045, 2009.

\bibitem{uddin2019towards}
Lucina~Q Uddin, BT~Yeo, and R~Nathan Spreng.
\newblock Towards a universal taxonomy of macro-scale functional human brain networks.
\newblock {\em Brain topography}, 32(6):926--942, 2019.

\bibitem{zeidan2022global}
Jinan Zeidan, Eric Fombonne, Julie Scorah, Alaa Ibrahim, Maureen~S Durkin, Shekhar Saxena, Afiqah Yusuf, Andy Shih, and Mayada Elsabbagh.
\newblock Global prevalence of autism: A systematic review update.
\newblock {\em Autism research}, 15(5):778--790, 2022.

\bibitem{penny2011statistical}
William~D Penny, Karl~J Friston, John~T Ashburner, Stefan~J Kiebel, and Thomas~E Nichols.
\newblock {\em Statistical parametric mapping: the analysis of functional brain images}.
\newblock Elsevier, 2011.

\bibitem{jenkinson2012fsl}
Mark Jenkinson, Christian~F Beckmann, Timothy~EJ Behrens, Mark~W Woolrich, and Stephen~M Smith.
\newblock Fsl.
\newblock {\em Neuroimage}, 62(2):782--790, 2012.

\end{thebibliography}

\appendix
\onecolumn

\section{Supplementary Materials}

\begin{table}[ht]
    \centering
    \begin{tabular}{c||c|c|c}
         &  ASD & HC & P-value\\
         \hline \hline
         Number & 417 & 483 & - \\
         \hline
         Sex (m/f) & 361/56 & 377/106 & 0.0117\\
         \hline
         Age (years) & 12.84 $\pm$ 5.04 & 13.84 $\pm$ 5.20 & 0.0017\\
         \hline
         PIQ & 105.78 $\pm$ 17.31 & 108.94 $\pm$ 14.54 & 0.0040\\
         \hline
         
         \end{tabular}
    \caption{Demographic and phenotypic information summarized on the group level, where age and PIQ are reported following the convention mean $\pm$ standard deviation, the group differences were tested using the Chi-Square test for sex, the Mann-Whitney U test for age, and Welch test for PIQ. Abbreviations: ASD, autism spectrum disorder; HC healthy controls; m/f, male/female; PIQ, performance intelligence quotient.}
    \label{tab:dems}
\end{table}

\begin{table}[ht]
    \centering
    \begin{tabular}{c|c}
        RSN & P-value \\
        \hline
        \hline
        Default Mode Network Anterior & 0.5158\\
        \hline
        Default Mode Network Posterior & 0.1654\\
        \hline
        Primary Visual Network & 0.3296\\
        \hline
        Lateral Visual Network & 0.7056\\
        \hline
        Salience Network & 0.3076\\
        \hline
        Auditory Network & 0.5048\\
        \hline
        Left Frontoparietal Network &  0.7996\\
        \hline
        Right Frontoparietal Network &  0.4128\\
        \hline
        Primary Sensorimotor Network & 0.4156\\
        \hline
        Lateral Sensorimotor Network & 0.6470\\
        \hline
        Cerebellum & 0.8012\\
        \hline
        Dorsal Attention Network & 0.1766\\
        \hline
        Language Network & 0.3550\\
        \hline
        Occipital Visual Network & 0.0782\\
        \hline
    \end{tabular}
    \caption{The results of two-sample t-statistic permutation testing using FSL's randomise. The reported P-values are the lowest voxel values (or the highest 1 - P-value, as this is how the results are returned) per resting-state network. This means there are no voxels closer to significance than the ones reported here. Note that P-values are corrected for multiple comparisons over voxels but not for multiple comparisons over RSNs, which would require further correction. Abbreviations: RSNs, resting-state networks. }
    \label{tab:permres}
\end{table}

\begin{figure}[ht]
    \centering
    \includegraphics[width=0.95\textwidth]{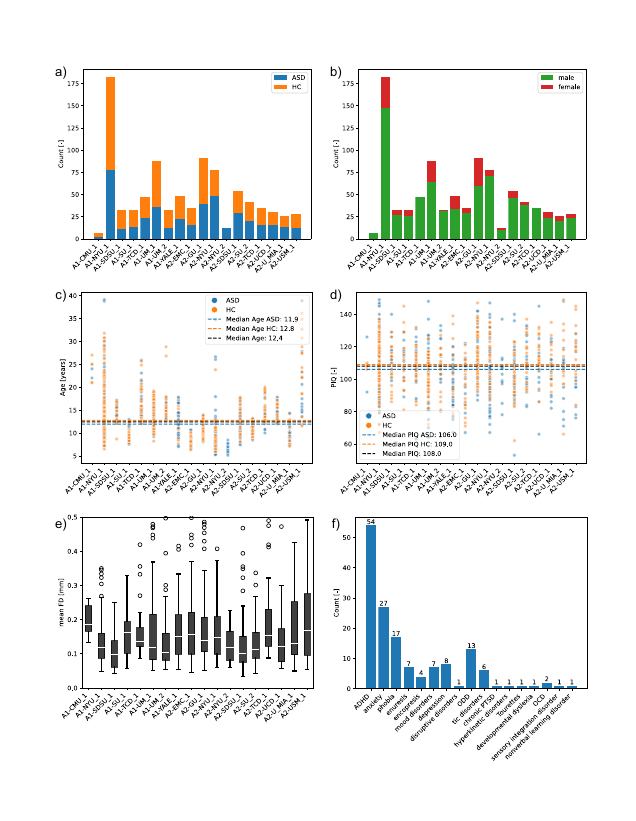}
    \caption{Information on phenotypes and motion. (a) Subjects with diagnostic labels per included site. (b) Subjects and sex distributions per included site. (c) The distribution of age and diagnosis per included site. (d) The distribution of PIQ and diagnosis per included site. (e) Mean framewise displacements per site. (f) Comorbidities present in the dataset. Abbreviations: ASD, autism spectrum disorder; HC, healthy controls; A1, ABIDE-I; A2, ABIDE-II; PIQ, performance intelligence quotient; FD, framewise displacement; CMU, Carnegie Mellon University; NYU, New York University; SDSU, San Diego State University; SU, Stanford University; TCD, Trinity College Dublin; UM, University of Michigan; EMC, Erasmus Medical Center; GU, Georgetown University; UCD, University of California Davis; UMIA, University of Miami; USM, University of Utah School of Medicine; ADHD, attention deficit/hyperactivity disorder; ODD, oppositional defiant disorder; PTSD, post-traumatic stress disorder; OCD, obsessive compulsive disorder.}
    \label{fig:demphen}
\end{figure}

\end{document}